\documentclass[%
reprint,                 
amsmath,amssymb,
aps,
]{revtex4-2}

\usepackage{graphicx}
\usepackage{dcolumn}
\usepackage{bm}

\usepackage{times}

\usepackage{comment}

\usepackage{hyperref}

\usepackage[bigfiles]{pdfbase}
\ExplSyntaxOn
\NewDocumentCommand\embedvideo{smm}{
	\group_begin:
	\leavevmode
	\tl_if_exist:cTF{file_\file_mdfive_hash:n{#3}}{
		\tl_set_eq:Nc\video{file_\file_mdfive_hash:n{#3}}
	}{
		\IfFileExists{#3}{}{\GenericError{}{File~`#3'~not~found}{}{}}
		\pbs_pdfobj:nnn{}{fstream}{{}{#3}}
		\pbs_pdfobj:nnn{}{dict}{
			/Type/Filespec/F~(#3)/UF~(#3)
			/EF~<</F~\pbs_pdflastobj:>>
		}
		\tl_set:Nx\video{\pbs_pdflastobj:}
		\tl_gset_eq:cN{file_\file_mdfive_hash:n{#3}}\video
	}
	\pbs_pdfobj:nnn{}{dict}{
		/Type/RichMediaInstance/Subtype/Video
		/Asset~\video
		/Params~<</FlashVars (
		source=#3&
		skin=SkinOverAllNoFullNoCaption.swf&
		skinAutoHide=true&
		skinBackgroundColor=0x5F5F5F&
		skinBackgroundAlpha=0.75
		)>>
	}
	\pbs_pdfobj:nnn{}{dict}{
		/Type/RichMediaConfiguration/Subtype/Video
		/Instances~[\pbs_pdflastobj:]
	}
	\pbs_pdfobj:nnn{}{dict}{
		/Type/RichMediaContent
		/Assets~<<
		/Names~[(#3)~\video]
		>>
		/Configurations~[\pbs_pdflastobj:]
	}
	\tl_set:Nx\rmcontent{\pbs_pdflastobj:}
	\pbs_pdfobj:nnn{}{dict}{
		/Activation~<<
		/Condition/\IfBooleanTF{#1}{PV}{XA}
		/Presentation~<</Style/Embedded>>
		>>
		/Deactivation~<</Condition/PI>>
	}
	\hbox_set:Nn\l_tmpa_box{#2}
	\tl_set:Nx\l_box_wd_tl{\dim_use:N\box_wd:N\l_tmpa_box}
	\tl_set:Nx\l_box_ht_tl{\dim_use:N\box_ht:N\l_tmpa_box}
	\tl_set:Nx\l_box_dp_tl{\dim_use:N\box_dp:N\l_tmpa_box}
	\pbs_pdfxform:nnnnn{1}{1}{}{}{\l_tmpa_box}
	\pbs_pdfannot:nnnn{\l_box_wd_tl}{\l_box_ht_tl}{\l_box_dp_tl}{
		/Subtype/RichMedia
		/BS~<</W~0/S/S>>
		/Contents~(embedded~video~file:#3)
		/NM~(rma:#3)
		/AP~<</N~\pbs_pdflastxform:>>
		/RichMediaSettings~\pbs_pdflastobj:
		/RichMediaContent~\rmcontent
	}
	\phantom{#2}
	\group_end:
}
\ExplSyntaxOff

\begin{document}
	
	\preprint{APS/123-QED}
	
	\title{Simulation of Matrix Product States to Unveil the Initial State Dependency of non-Gaussian Dynamics of Kerr Nonlinearity}
	
	\author{Souvik Agasti}
	\email{souvik.agasti@uhasselt.be}
	
	\affiliation{
		IMOMEC division, IMEC, Wetenschapspark 1, B-3590 Diepenbeek, Belgium
	}%
	\affiliation{
		Institute for Materials Research (IMO), Hasselt University,	Wetenschapspark 1, B-3590 Diepenbeek, Belgium
	}%

	
	\begin{abstract}
		We simulate a free dissipative and coherent-driven Kerr nonlinear system using a time-evolving block decimation (TEBD) algorithm, to study the impact of the initial state on the exact 
		{quantum} dynamics of the system. The superposition of two coherent branches results in non-classical time dynamics. 
		The Wigner state representation confirms that the system ends up saturating to two different branches, through evolving different trajectories, resulting in de-Gaussification throughout evolution.
		Furthermore, we also see that the time evolution suffers a residual effect of the initial state.
		
		.		
	\end{abstract}
	
	\keywords{Kerr nonlinearity, Wigner function, Density matrix renormalization group algorithm, non-Gaussianity, Open quantum system}
	\maketitle
	

	\section{INTRODUCTION}



	The Kerr effect has shown intensity-dependent phase shift of
	light during propagation due to non-linear quadratic electro-optic (QEO) response which has remained in interest over centuries, for providing a broad range of applications in many optical and magnetic devices. 
	The Kerr nonlinearity has shown some interesting quantum phenomena such as photon bunching and antibunching \cite{Kerr_Drummond_Walls}, photon switching in quantum interference \cite{Photon_Switching}, dynamical optical bistability via bifurcation process \cite{Yasser_Sharaby_bifurcation} and the generation of non-classical states \cite{Agasti_kerr}. 
	All the setups exhibit bistability due to nonlinear susceptibility as their predominant characteristic.	
	Theoretically, the multistability of a steady-state Kerr nonlinear system is treated in two different ways. One way of doing it is by approximating the state to a coherent one and treating it semi-classically using the quantum Langevin equation \cite{McCall1974kerr, Agrawal_Carmichael_kerr}. On the other hand, it is done by mapping the master equation to the Fokker-Planck equation which determines the exact quantum mechanical solution  \cite{Walls_Carmichael_kerr1977, Kerr_Drummond_Walls}. 
	While implementing both theoretical techniques for the analysis of the steady state, the evolution of an anharmonic single-mode field with Kerr nonlinearity is generally considered, interacting with a zero-temperature reservoir under coherent driving.
	
	{
		In this context, it is essential to mention that such nonlinearities are involved with quantum-limited amplification and parametric coupling, observed in superconducting microwave cavities \cite{superconducting_cavity_yale_1}, which has also been used for quantum control and manipulation \cite{ superconducting_cavity_yale_2}.
	} 
	Besides, Kerr nonlinearity 
	has played a crucial role	in the generation of squeezed states \cite{30_years_squeezed_light_generation, 75_years_squeezed_light_generation}, particularly using optical fibers \cite{Kerr_Effect_Fibers}. Recently, squeezing based on third-order nonlinear susceptibility has also been used for the increment of sensitivity in interferometers \cite{Kerr3_sqz}. The propagation of light in Kerr-based fibers requires long propagation distances and high powers since the Kerr nonlinearity exhibits to be very small in silica glass \cite{Boyd_kerr_optics_silica_glass}. 	In this limit, a standard classical approach based on Maxwell’s equations has been used which leads to a set of coupled nonlinear Schr\"odinger equations \cite{nonlinear_Schrodinger_equations}, which describes the behavior of optical fields in nonlinear dispersive media which has been seen diversely in the phenomena e.g. self-focusing of ultrashort pulses \cite{Self_focusing_kerr_theory, Self_focusing_kerr_Experimental, Agasti_OAM_tampere}, pulse generation \cite{pulse_generation} and fiber solitons  \cite{fiber_solitons}. However, there are interesting unavoidable quantum effects that cannot be visible	in this classical approach. Therefore, the borderline between classical and quantum descriptions becomes a genuine question to think about. We explore this conundrum here by investigating Wigner's function in
	phase-space methods \cite{Quantum_optics_Walls_Milburn}, and quantumness of the state through fidelity measurement with the possible nearest approximated classical state \cite{Quantum_fidelity_measurement}.
	Another important aspect that we explore here is the impact of the initial state on the dynamical behavior of such nonlinear quantum systems surviving under constant drive and spontaneous decoherence, which is experienced previously in different systems, e.g. impurity-infected solid-state systems \citep{myprevious_prl}, and even in Kerr nonlinear systems \cite{my_physica_scripta}. 

	
	{
		The generation of quantum state in a driven nonlinear dissipative system has proven to be a useful tool in quantum information processing.  
		Such quantum-optical treatment of driven and dissipative oscillators with Kerr nonlinearity is important in the context of resonators, especially micro and nano resonators such as photonic crystals \cite{photonic_crystal} and waveguides \cite{waveguide}. The description of lossy propagation in fibers and crystals, and switching between two branches shows hysteresis which has recently been observed experimentally \cite{hysteresis_kerr_experimental_josa, hysteresis_kerr_experimental_PRL} and numerically \cite{hysteresis_kerr_theory_PRA, kerr_time_dynamics,  my_physica_scripta} in Kerr nonlinear system. However, those approaches remained unable to reveal the nonlinearity-induced mixed-state solutions of the quantum fluctuations and dynamical trajectories. 
		Even though efforts had been made to develop a quantum mechanical exact analytical solution for the dissipative anharmonic-oscillator by solving Liouville equation \cite{Dissipative_Quantum_Classical_Liouville_1, Dissipative_Quantum_Classical_Liouville_2}, it does not show the effects of the presence of external drive.		
		Therefore, the demand for studying the exact dynamical behavior of the state under coherent drive is still there, and how they are being affected by the lifetime of the system and their initial state; which is the key focus of this article.
	}

	The theoretical investigation of quantum dynamics of Kerr nonlinear systems is dependent on the Markovian dynamics of open quantum systems, which is often done by linearized approximation. 
	This often leads to overlooking non-linearity introduced interesting dynamical quantum effects in Kerr systems. Especially, one can not determine the exact dynamical behavior and the impact of the initial state, analytically.
	Besides, the existing theoretical framework does not explain the dependency of a steady state on the initial state of the system.  
	This limitation of analytics motivates numerical treatment to deal with the time evolution. 
	For this purpose, we use time-evolving block decimation (TEBD) algorithm which is also capable of time evolution simulations of time-dependent Hamiltonians.
	To do that, we transform the multimode environment to a one-dimensional chain \cite{Agasti_TEBD_conference} which is composed of numerical diagonalization and renormalization process  \citep{Agasti_kerr, Agasti_OQS}. 
	{The algorithm demands to express the state of the chain as matrix product state (MPS) and the time evolution goes through simulating the lattice chain in alternative pairs \cite{Vidal_TEBD}. }

	{
		In this article, besides analytical theory, we simulate a coherent-driven Kerr nonlinear system using the TEBD algorithm, to investigate the impact of the initial state on its exact dynamics. We initiate here with the consistency of the numerical results to its corresponding analytical counterpart, along with the quantum jump of the system and how the bistable nature of the steady state is dependent on the external drive.  Earlier, in \cite{Agasti_kerr, my_physica_scripta}, while simulating the Kerr nonlinear system using TEBD, we studied the coherent dynamics of the system and the quantum fluctuations associated with it. Unlike that, however, this article investigates the exact quantum dynamics of the state for the first time and the influence of the initial state on it. 	
		We visualize it through the evolution of the Wigner function which gives a picture of how the state settles on the different locations of phase space. We then characterize the quantumness of the dynamics by studying the degree of coherence and fidelity. For better understanding, we furthermore study the non-Gaussian nature of the evolution. 
		The investigation gives a better understanding of the evolution of the system, which provides an intuitive range of acceptance in which classical or coherent approximation remains valid.
	}

	\begin{figure}[t!]
		\includegraphics[width=1\linewidth]{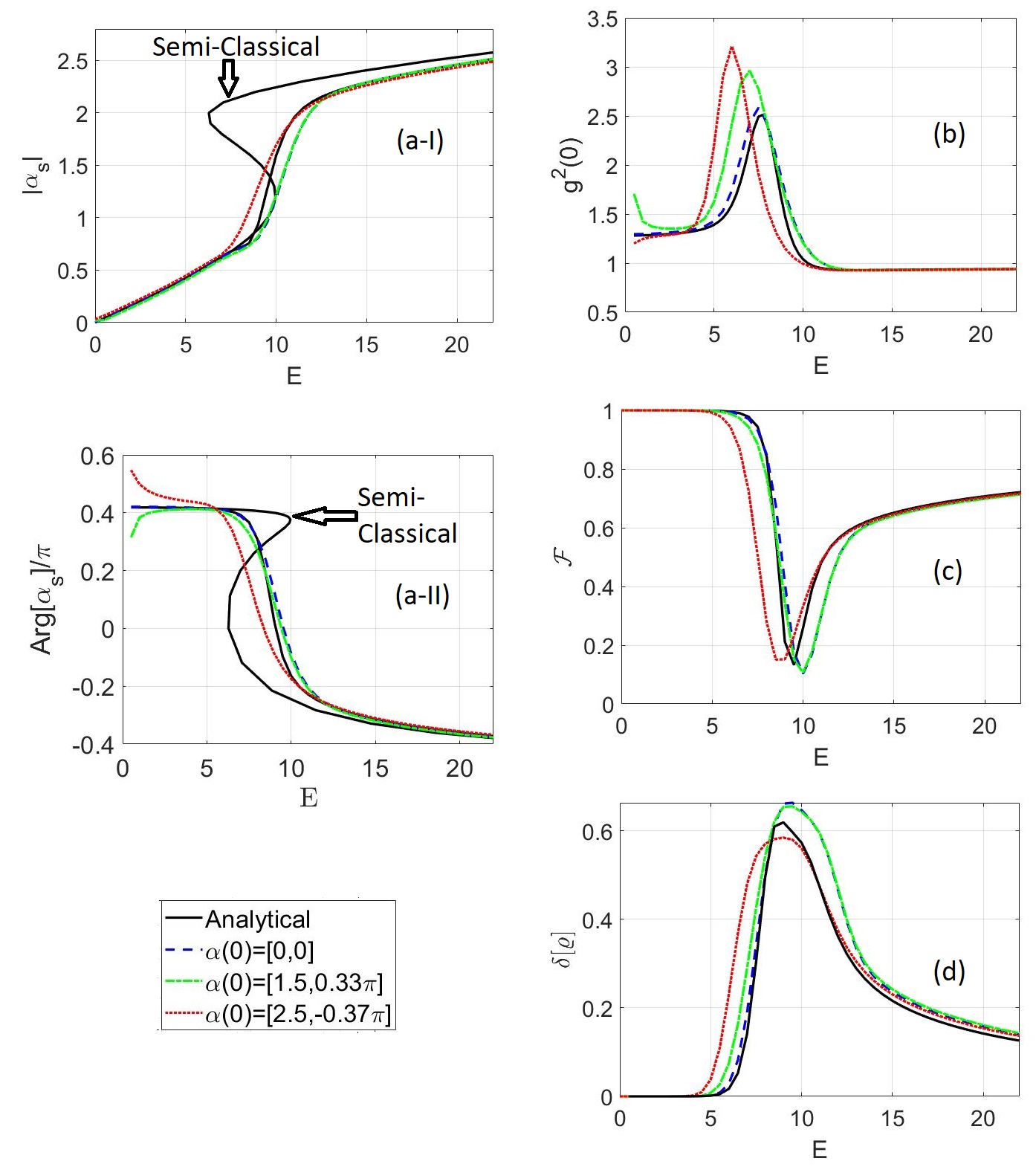}
		\caption{ Steady state (a-I) Field amplitude and (a-II) phase, (b) second-order correlation function, (c) fidelity and (d) non-Gaussianity of the Kerr-nonlinear system with the variation of driving field amplitude for $\Delta=-12 g,\chi"= 1.5 g, \gamma= 6.28 g$. TEBD simulation parameters $N = 61, x_{max} = 60, \chi = 36, M = 20, \delta t = 10^{-2} g^{-1}$ and total time of evolution $2 g^{-1}$. }
		\label{fig:kerr_plot}
	\end{figure}

	\begin{figure*}[t!]
		\includegraphics[width=1\linewidth]{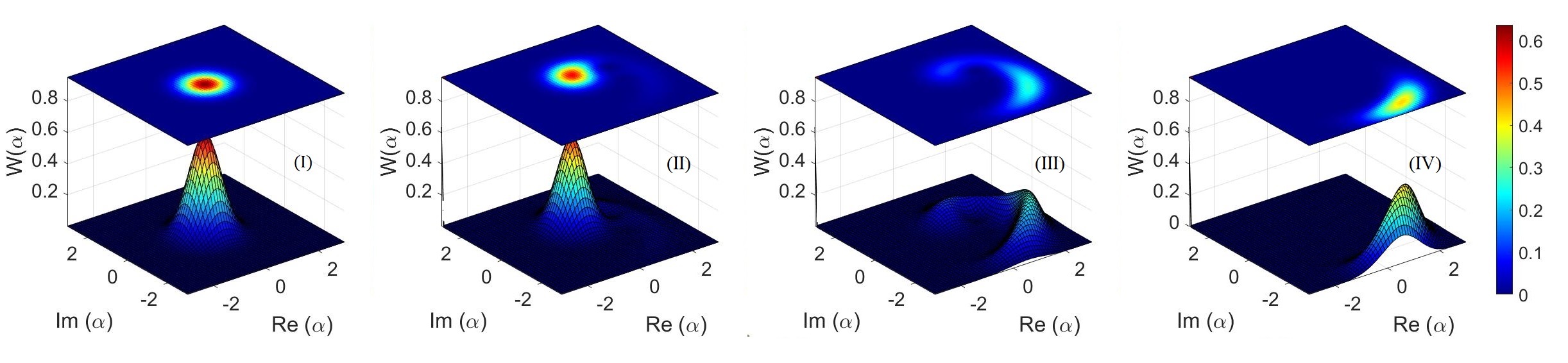}
		\caption{ Analytically determined steady-state Wigner function for (I) E$= 1$, (II) E$=8$, (III) E$=10$ and (IV) E$= 20$
			. All other parameters remain same with Fig. \ref{fig:kerr_plot}. }
		\label{fig:wig_fn_anal_E1_8_10_20}
	\end{figure*}

	\begin{figure*}[t!]
		\includegraphics[width=1\linewidth]{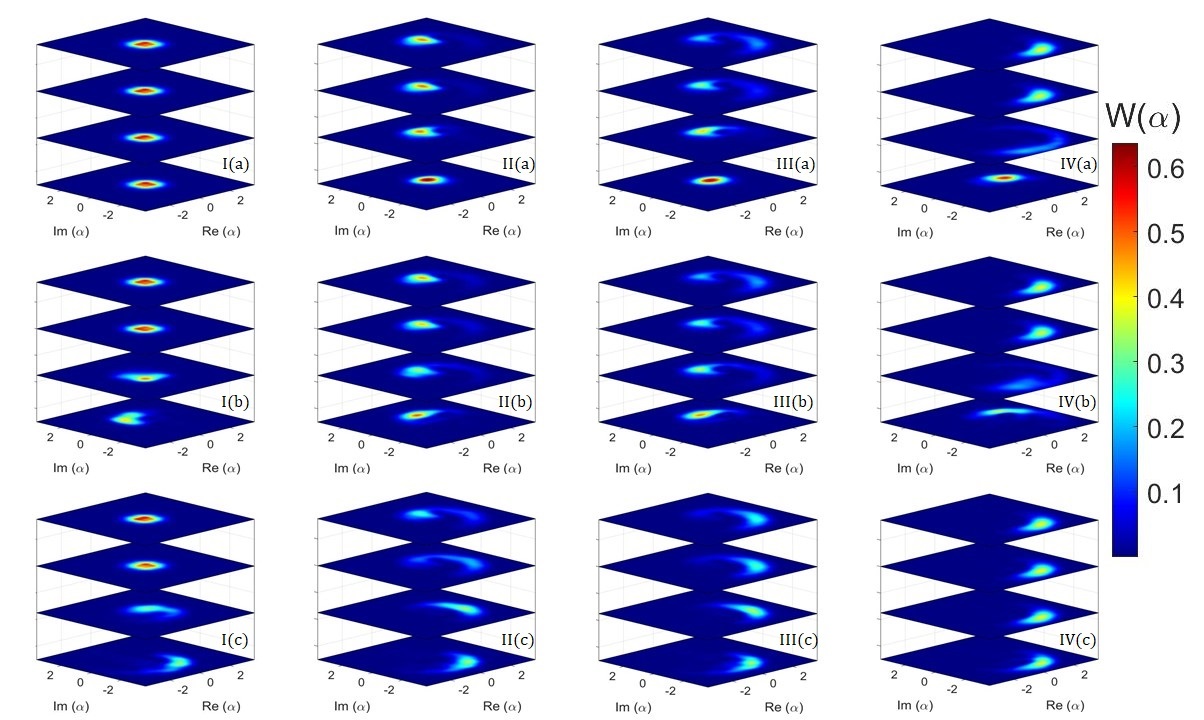}
		\caption{  Time evolution of Wigner function for (I) E$= 1$, (II) E$=8$, (III) E$=10$ and (IV) E$= 20$, and initial field (a)$\alpha(0)=[0,0]$, (b)$\alpha(0)=[1.5,0.33\pi]$ and (c)$\alpha(0)=[2.5,-0.37\pi]$
			; for the time $0.1, 0.3, 0.8, 2 g^{-1}$ (from bottom to top)	. All other parameters remain the same with Fig. \ref{fig:kerr_plot}.
			See videographic representation of the evolution of Wigner function in  \href{supplimentary.pdf}{Supplementary Material} .
		}
		\label{fig:wig_fn_1_6_16_41_101}
	\end{figure*}

	\begin{figure*}[t!]
		\includegraphics[width=1\linewidth]{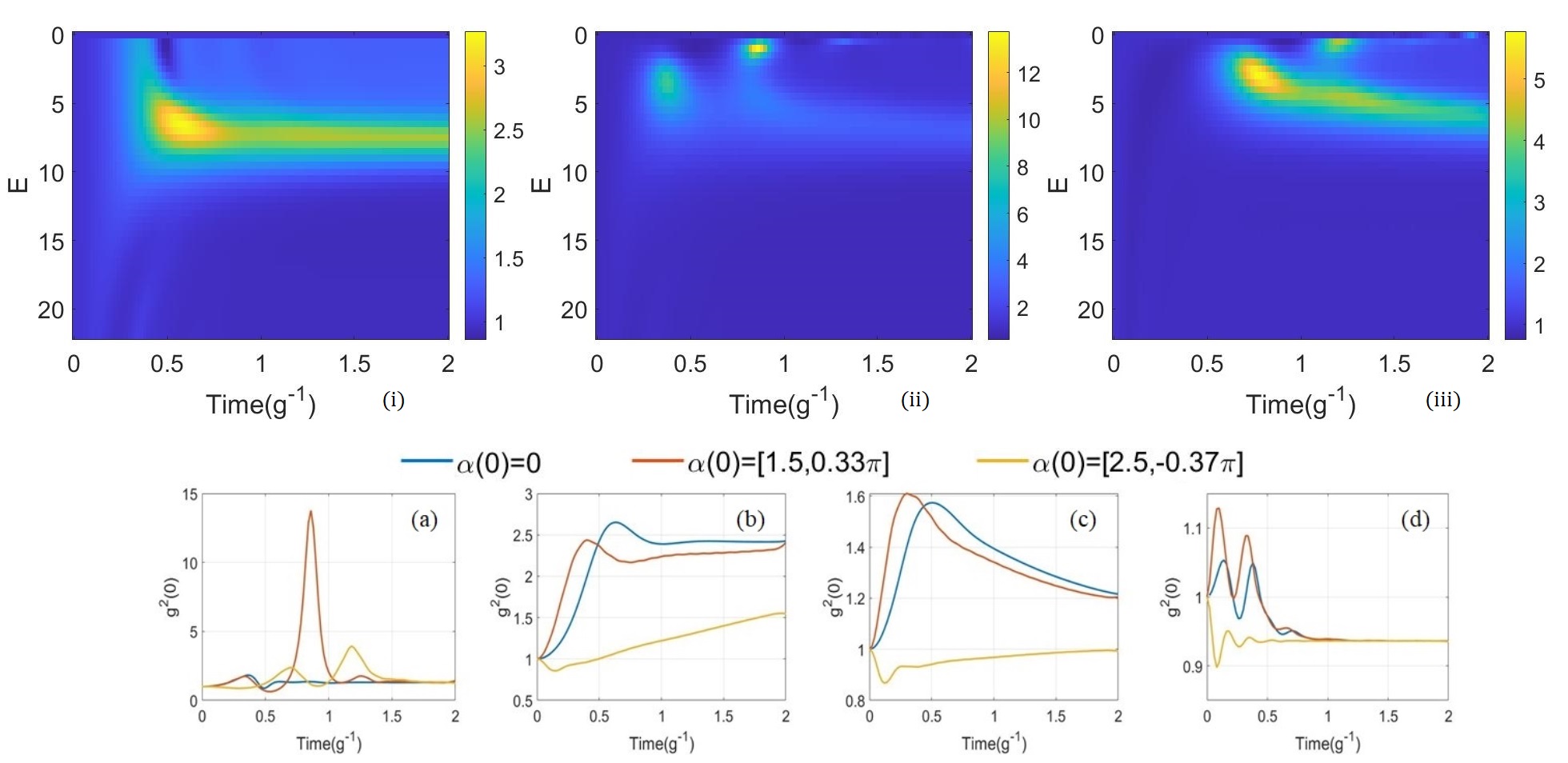}
		\caption{Evolution of second order correlation function for the initial field (i)$\alpha(0)=[0,0]$, (ii)$\alpha(0)=[1.5,0.33\pi]$ and (iii)$\alpha(0)=[2.5,-0.37\pi]$ in the top row, and 			
			(a) E$= 1$, (b) E$=8$, (c) E$=10$ and (d) E$= 20$ in bottom row 
			. All other parameters remain same with Fig. \ref{fig:kerr_plot}.}
		\label{fig:g20}
	\end{figure*}

	\begin{figure*}[t!]
		\includegraphics[width=1\linewidth]{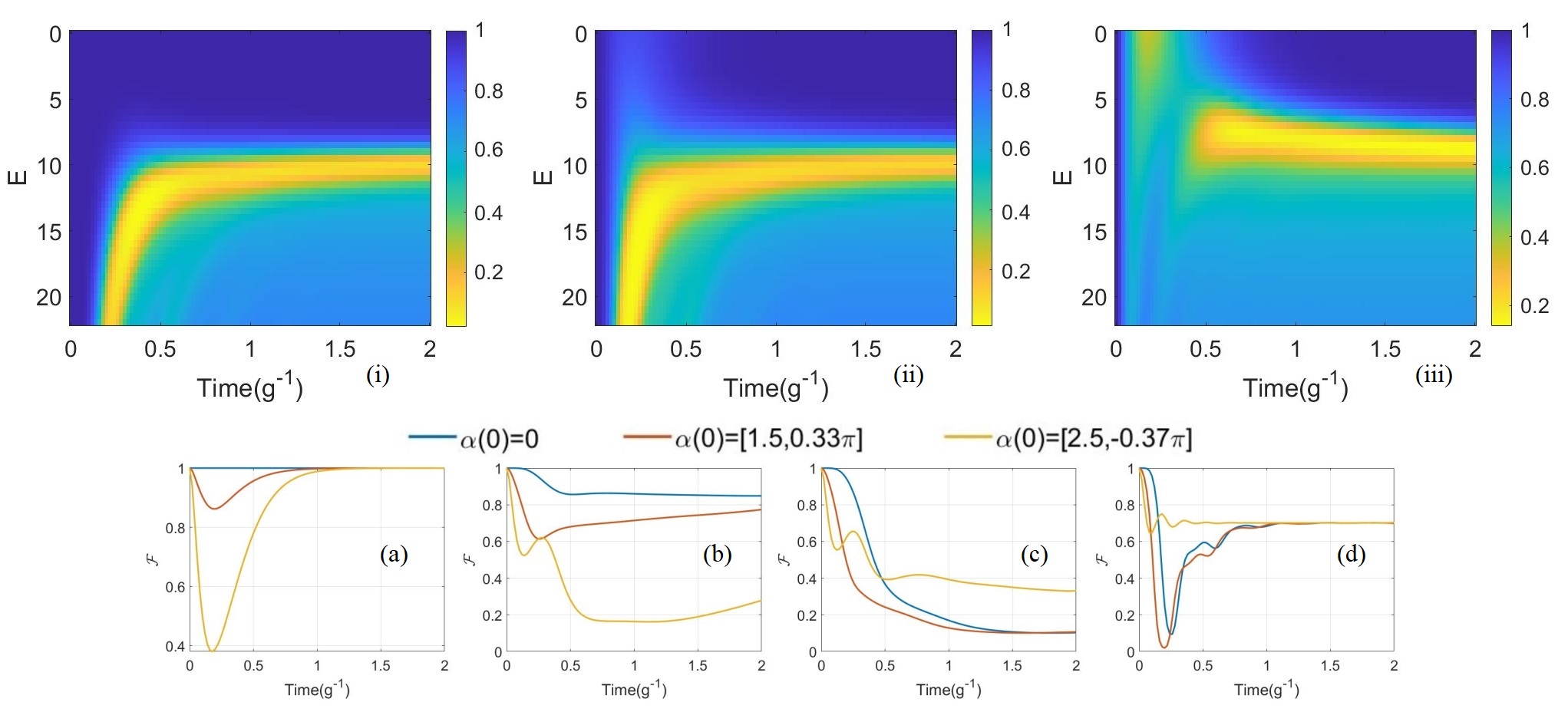}
		\caption{Fidality of the states while evolving for the initial field (i)$\alpha(0)=[0,0]$, (ii)$\alpha(0)=[1.5,0.33\pi]$ and (iii)$\alpha(0)=[2.5,-0.37\pi]$ in the top row, and 			
			(a) E$= 1$, (b) E$=8$, (c) E$=10$ and (d) E$= 20$ in bottom row  
			. All other parameters remain same with Fig. \ref{fig:kerr_plot}.}
		\label{fig:Fidality_E1_8_10_20}
	\end{figure*}

	\begin{figure*}[t!]
		\includegraphics[width=1\linewidth]{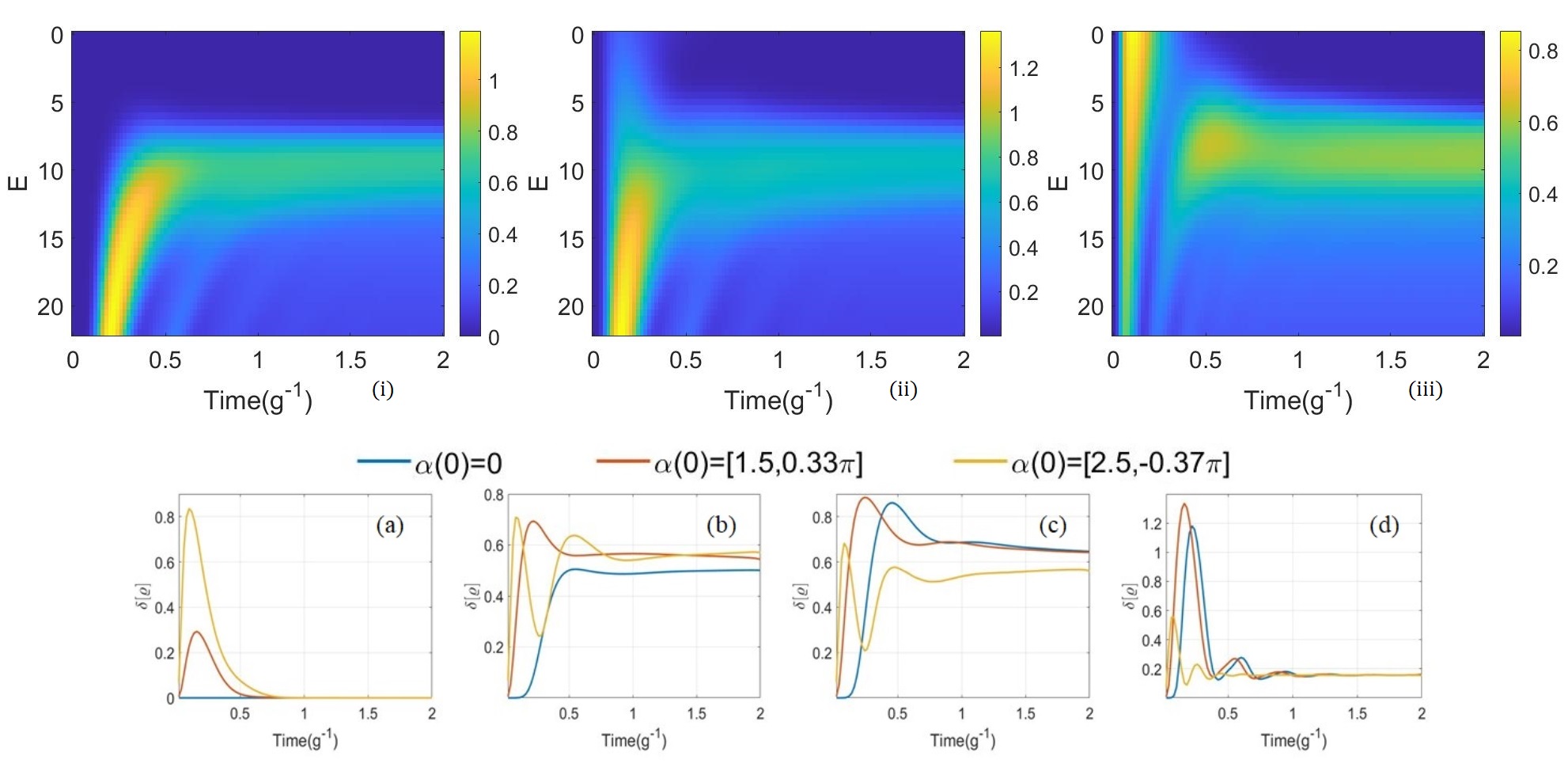}
		\caption{ Non-Gaussianity of the states while evolving for the initial field (i)$\alpha(0)=[0,0]$, (ii)$\alpha(0)=[1.5,0.33\pi]$ and (iii)$\alpha(0)=[2.5,-0.37\pi]$ in the top row, and 			
			(a) E$= 1$, (b) E$=8$, (c) E$=10$ and (d) E$= 20$ in bottom row  
			. All other parameters remain same with Fig. \ref{fig:kerr_plot}.}
		\label{fig:nonG_E1_8_10_20}
	\end{figure*}

	\section{THE SYSTEM}\label{sec:level2}
	
	The Hamiltonian of a Kerr nonlinear system with an external drive is given by
	
	\begin{equation}\label{kerr_hamiltonian}
		{H}_S =\omega_S a^\dagger a+\chi"{a^\dagger}^2a^2+i(a^\dagger Ee^{-{i\omega _{L}t}}- a E^*e^{{i\omega _{L}t}})
	\end{equation}
	
	where $a^\dagger$ and $a$ are the creation and annihilation operators of the system. The oscillation frequency of the mode of the system is $\omega_S$. $\chi"$ is the anharmonicity parameter which is generated from the real part of the third-order nonlinear susceptibility tensor.  The external driving field acting on the system  $\vec{\tilde{E}}(t)=\vec{E}e^{-i\omega_L t}+\vec{E}^* e^{i\omega_L t}$ has an amplitude  $E$ and oscillation frequency $\omega_L$. Moving to the frame of the driving field gives the cavity frequency detuned to $\Delta=\omega_S-\omega_L$. Introducing a dissipative thermal bath (B) with the system (S), the total Hamiltonian gives
	
	\begin{equation}
		H_{tot} = H_S + H_B + H_{SB} 
	\end{equation} 
	
	where $H_B =  \lim_{x_{m}\to \infty} \int_{-x_{m}}^{x_{m}} g(x) d^\dagger (x) d(x) \mathrm{d}x $ represents the Hamiltonian of a multimode thermal reservoir which is expected to be at zero temperature, and $H_{SB} =  \lim_{x_{m}\to \infty} \int_{-x_{m}}^{x_{m}} h(x)\left(a^\dagger d(x) + h.c.\right) \mathrm{d}x $ is the interaction Hamiltonian.  $g(x)$ is the frequency of oscillation and for the environmental mode $x$.  $h(x)$ is the coupling strength between the system and environment around the mode of oscillation of the system $\omega_S$. $d^\dagger_x$ and $d_x$ are the creation and annihilation operators that satisfy the usual bosonic commutation relation $[d^\dagger_x, d_{x'}] = \delta_{x,x'}$. Here, we consider a linear dispersion relation: $g(x) = g. x,$ for the conventional system/bath (S/B) couplings, where $g$ is the inverse of the density of states. In this process, we have chosen a hard cutoff frequency limit of the bath: $\omega_c = g.x_{m}$. In addition, the consideration of Markovian dynamics within the range of frequencies of interest makes the S/B coupling strength mode independent (wide band limit approximation):  $h(x) =c_0 $ \cite{book_robert}, which characterizes the properties of the bath through a well-defined spectral density function $J(\omega)$  \cite{spectral_density_function}, as
	
	\begin{equation}
		\label{spctral_density}
		J(\omega) = \frac{1}{2} \gamma\Theta(\omega+\omega_c)\Theta(\omega_c-\omega),
	\end{equation} 
	
	where $\gamma = 2\pi c_0^2$ is the rate of dissipation of the system and $\Theta$ is the Heaviside step function. 
	The Heisenberg-Langevin equation of motion  can be obtained semi-classically from the system Hamiltonian in Eq. \eqref{kerr_hamiltonian}, which in turn determines the stationary system field, as
	
	\begin{align}\label{kerr coherent solution}
		|E|^2=|\alpha|^2\left((\Delta+2\chi"|\alpha|^2)^2+\frac{\gamma^2}{4}\right),
	\end{align}

	where $\alpha$ is the steady-state system field. 
	{	The semi-classical analytical solution determines the branch values and the transition region \cite{my_physica_scripta}. In order to observe classical bistability, the detuning has to be adjusted to $\Delta < -\gamma \sqrt{ \frac{3}{4} }$.
		Based on the theoretical framework defined here, we simulate the dynamical behavior numerically, which includes the transformation of the S/B coupling model to a 1D chain and simulates afterward using the TEBD algorithm  (See Appendix \ref{TEBD_NUMERICAL_MODEL})\cite{Agasti_OQS}. Following the algorithm, we express the state of the chain as MPS using Schmidt decomposition and perform real-time simulation using second-order Suzuki-Trotter expansion \cite{Vidal_TEBD}. 	 The technique has been used successfully before in \cite{my_physica_scripta, Agasti_kerr} for the simulation of the Kerr nonlinear system. We also explained the applicability and limitations of the technique in \cite{Agasti_kerr}, where we explained numerical complexity and the reliable time steps that can be accepted for the simulation. Besides, we also explained the reliable size of the chain, i.e. number of sites, the size of Hilbert space, and the Schmidt number.
	}

	The Kerr nonlinear bistability, seen in \ref{kerr coherent solution}, motivates us to consider three different initial conditions, to investigate the switching effects on the dynamical behavior. Earlier in \cite{my_physica_scripta}, we have seen that the dynamical behavior and the operational frequency are highly dependent on the initial state of the system. 

	\section{ STEADY STATE} 
	
	In Schr\"odinger picture, one can derive the exact quantum mechanical solution of the moment calculating generalized function by mapping the master equation into the Fokker-Planck equation \cite{Kerr_Drummond_Walls}. This in turn helps to determine the steady-state field amplitude ($\alpha_{S}$), second-order correlation function ($	g^2(0) $) for the coherent field $\alpha$ as
	
	\begin{subequations}
		\label{Kerr master g20 and field}
		\begin{align}
			\alpha_{S} &= G^{(0,1)} 
			\label{Kerr master field} \\
			g^2(0) &= G^{(2,2)}/ (G^{(1,1)}) ^2 
			\label{subeq:g20}
		\end{align}
	\end{subequations}
	
	where
	
	\begin{widetext}
		\begin{equation}\label{moment_calculating_generalized_function}
			G^{(m,n)} = \frac{(-1)^n (E/i\chi")^{(m+n)} \Gamma(p) \Gamma(q) F\big(p+n,q+m,2[E_0/\chi"]^2\big) }{ \Gamma(p+n) \Gamma(q+m) F\big(p,q,2[E_0/\chi"]^2\big) } 
		\end{equation}
	\end{widetext}
	
	is the moment calculating a generalized function. Here $E_0=E=E^*, p=[\frac{\Delta}{\chi"}+\frac{\gamma}{2i\chi"}],q=[\frac{\Delta}{\chi"}-\frac{\gamma}{2i\chi"}]$ and $F\big(p,q,z\big)=F\big([],[p,q],z\big)$ is the $_0F_2$ hypergeometric function. 
	
	
	We compare analytically determined exact quantum mechanical steady-state behavior from Eq. \eqref{Kerr master g20 and field} in Fig. \ref{fig:kerr_plot} to compare with the corresponding numerical counterpart determined by the TEBD numerical methods. The comparison justifies the consistency between them. 
	The amplitudes and phases of the steady-state system fields (Eq. \eqref{Kerr master field}) are plotted in Fig. \ref{fig:kerr_plot} (a)(I and II, respectively); which shows, both the solutions, moreover, follows each other. Also, the solutions have been seen to remain analogous to classical bistability. However, unlike semi-classical solutions, the quantum mechanical solution  gives a superposition of two stable branches and, therefore, does not exhibit 
	{multiple stable states for any particular drive}. Following that, the steady-state system field loses its linear behavior with the increment of the external drive, and the system tends to jump from one steady state to another. As the coherent states are non-orthogonal, we see a non-classical states are generated in the quantum mechanical solution, especially around the transition region. 
	{ Therefore, we observe a peak in Fig. \ref{fig:kerr_plot} (b), in the plot of second-order correlation function $(g^2(0))$ at that region. The non-classical nature of the evolution is confirmed by the deviation of $(g^2(0))$ from the unit value, which can be understood by the Eq. \eqref{subeq:g20}. For weaker drives the quantity is seen to be $(g^2(0)) > 1$ which is caused by the bunching modes of the super-Poissonian statistical distribution of photons (variance $>$ average photon number). On the other hand, the system is seen to stabilize to an anti-bunching mode with the sub-Poissonian distribution of photons (variance $<$ average photon number), for stronger drives; which may lead to obtain squeezed states. Therefore, evolution in two branches is anticipated to exhibit opposite characteristics. 
		The plot also indicates that the moving from lower to the upper branch while increasing the strength of the drive, causes the system to undergo from bunching to the anti-bunching steady-state mode. 
	}

	Another quantitative way to visualize how close the quantum state is to its quasiclassical solution is the corresponding fidelity \cite{Quantum_fidelity_measurement}. The quantity is expressed as
	
	\begin{equation}
		\mathcal{F}(t) = \text{ Tr} \left[ \varrho(t) \varrho_{cl}(t)\right] = \pi \int \mathrm{d} \alpha W(\alpha) W_{cl} (\alpha)
	\end{equation}
	
	which measures the overlap of the quantum state $(\varrho)$ to its nearest classical one $(\varrho_{cl})$ 
	{ which is defined by $\varrho_{cl}(t) = \vert \alpha_{cl}\rangle\langle \alpha_{cl}\vert$, where $\alpha_{cl}(t) = \text{Tr}\left[ a\varrho(t)\right]$}. Another way, one can say it is a measure of the overlap between steady-state Wigner function ($W(\alpha)$) to 
	that of the closest classical one  
	{ $W_{cl}$ representing Wigner function of the state corresponds to the state $\varrho_{cl}(t)$}, which is expected to be perfectly Gaussian in shape. Following the measurement of  $(g^2(0))$, the fidelity measurement in Fig. \ref{fig:kerr_plot}(c) shows that the steady state deviates the most from its closest classical counterpart around the transition region. However, the fidelity remains
	larger for low-intensity fields in the lower branch compared to the upper one. 
	{Since the result is a prescription for the mathematical quantification of the degree of similarity between the quantum state to its closest classical one, it signifies the study of quantum dynamics through linearized approximation (as done previously in \cite{my_physica_scripta}) gives less error when the state undergoes thorough one among two branches (especially, at lower branch). The quantumness of the state becomes more robust around the transition region.  Since the fidelity quantifies how close is the produced state to its classical counterpart, the information is useful in quantum information processing, as the preparation of a quantum state is limited by imperfection}

	The non-classicality of the system provokes to check the non-Gaussian nature of the system, which has been of interest over decades for its application in quantum communication \cite{PRL_kerr_quantum_comm, PRA_kerr_quantum_comm} and 
	optomechanical systems \cite{optomechanics_nonG_Xuereb}. The non-Gaussianity of a continuous variable state $(\varrho)$ is typically quantified by the quantum relative entropy as 
	
	\begin{equation}\label{non_G_measure}
		S(\varrho||\varrho_G) = \text{Tr}\left[ \varrho (\log \varrho - \log \varrho_G)\right]
	\end{equation}
	
	measuring distance between the quantum state $\varrho$ and the closest Gaussian state $\varrho_G$ as a reference $(0 \le S(\varrho ||\varrho_G) \le \infty)$ \cite{nonG_Banaszek_1, nonG_Banaszek_2}, where $S(\varrho) = -\text{Tr}\left[ \varrho (\log \varrho)\right]$ is the von Neumann Entropy. The closest Gaussian state $\varrho_G$ is defined to have the same first and second moments, and therefore, the same correlation matrix. The plot of non-Gaussianity in Fig. \ref{fig:kerr_plot}(d) shows that the deviation from Gaussianity maximizes when the system goes into the transition region, and the lower branch stabilizes closer to a Gaussian state compared to the upper one.

	The non-classical steady-state can better be understood from the shape of the Wigner function, i.e. deviation from its Gaussian nature. Even though, the dynamics of the state evolving freely under the influence of Kerr nonlinearity has been studied before through the Wigner function \cite{free_kerr_wigner}, the dissipation was not considered. Therefore, the dynamics of the dissipative Kerr nonlinear system are yet to be studied. 	
	The steady-state Wigner function ($W(\alpha)$) can be calculated analytically from the solution of the master equation using moment calculating generalized function (Eq. \eqref{moment_calculating_generalized_function}) (See Appendix \ref{Analytical_Representation}):
	
	\begin{equation}
		W(\alpha) = \frac{2}{\pi}  e^{-2|\alpha|^2 } \sum_{k,m,n} \frac{(-1)^k 2^{k+m+n} }{k!m!n!} {\alpha^*}^n \alpha^m G^{(k+m,k+n)}  . \label{Wigner_function}
	\end{equation}
	
	Fig.  \ref{fig:wig_fn_anal_E1_8_10_20} plots the Wigner function from the analytical steady state solution of Eq. \eqref{Wigner_function}, which exhibits, while increasing the strength of the drive (I-IV), the bump is seen to be stabilized at different locations of phase space at a steady state. The bump is seen to be splitting into two around the transition region (II and III), which reminds the fact that the state is constructed by the superposition of two coherent states belonging to two different branches. Eventually, the shape loses its Gaussian nature and generates a non-classical state.

	The initial state has been shown to have a clear impact on the steady-state behavior of the system. 
	For different initial states, a residual effect is observed in the quantum jumps from one branch to another, which occurs at the different driving fields around the classically determined transition region. Fig. \ref{fig:kerr_plot} (a) shows that the steady state intends to transit from bunching to anti-bunching mode earlier in the TEBD-determined quantum mechanical estimation, for the initial state belonging in the upper branch. This results shifting the peak of $(g^2(0))$ in Fig. \ref{fig:kerr_plot} (b), towards the weaker drive. 
	Therefore, this also causes to push the dip towards a weaker field when we estimate fidelity numerically in Fig.  \ref{fig:kerr_plot} (c). 
	Following non-classicality, the non-Gaussianity measurement has also witnessed a similar phenomenon in Fig. \ref{fig:kerr_plot}(d), where we observe the bump of non-Gaussianity moves toward the lower driving field, around the classical transition region, when the system starts evolving from a stable state of higher branch.
	The phenomenon can also be understood better by comparing the numerically determined Wigner function to its analytical counterpart. The steady-state Wigner function in the top plots of Fig. \ref{fig:wig_fn_1_6_16_41_101}
	(a-d)(also in \href{supplimentary.pdf}{Supplementary Material}) shows the inevitable difference with Fig. \ref{fig:wig_fn_anal_E1_8_10_20}, as the bump moving earlier when the system starts evolving from an initial state at the upper branch, causing the early shift of the peak(dip) of non-Gaussianity (fidelity) in Fig. \ref{fig:kerr_plot}. It, therefore, inspires to study the exact time dynamics of the Kerr nonlinear system to visualize the evolution of the state.
	
	It is to be noted that analytically determined results of steady states do not show any impact on the initial state of the system. This happens because it presents an ideal result that can be obtained after an infinite time of evolution. Therefore, it can be anticipated that the increment of the time of evolution can reduce the hysteresis effect, directing the steady state narrowing down toward analytical results. Intensive attempts are made to understand such hysteresis in Kerr nonlinear systems experimentally in \cite{hysteresis_kerr_experimental_josa, hysteresis_kerr_experimental_PRL} and numerically in \cite{hysteresis_kerr_theory_PRA, my_physica_scripta, kerr_time_dynamics}.

	\section{QUANTUM DYNAMICS}

	The exact dynamical behavior and the evolution of the state can better be understood through the evolution of the Wigner function in Fig. 
	\ref{fig:wig_fn_1_6_16_41_101}
	, representing the quasi-probabilities in phase space for different times. Since the initial states are coherent, it determines the phase space location of the Wigner functions which is shaped to be a Gaussian hump. 
	Following the semi-classically determined upper branch, under stronger drive, the bump of the Wigner function stabilizes rotating clockwise rotation ( Fig. \ref{fig:wig_fn_1_6_16_41_101}
	IV(a) and IV(b)), whereas the bump rotates in a counter-clockwise direction when the system falls to a lower branch under a weaker drive ( Fig. \ref{fig:wig_fn_1_6_16_41_101}
	I(b) and I(c)). 
	As the photons are distributed oppositely in two different branches (super-Poissonian in the lower and sub-Poissonian in the upper branch), the dynamical behavior is anticipated to be opposite, which we discussed in the previous section. 
	{The rotational direction is determined by the branch values (including their phases) of the internal field \cite{kerr_time_dynamics, Agasti_kerr}.}
	{The numerical method also remains helpful to provide a parallel description of non-driven dissipative nonlinear systems, which one can compare with analytics given in \cite{Dissipative_Quantum_Classical_Liouville_1, Dissipative_Quantum_Classical_Liouville_2}
	}
	The phenomenon is also hinted in \cite{my_physica_scripta} where we determined the dynamical behavior of the classical field of the different branches of the system. 
	{However, this article presents a way to visualize the evolution of state, by studying its exact quantum dynamics.}
	Interestingly, in the transition region, the bump splits into two and moves in opposite directions to stabilize two different branches ( Fig. \ref{fig:wig_fn_1_6_16_41_101}
	(II and III)(a-c)). The existence of both the bumps together exhibits the superposition of two different coherent states, resulting in a non-classical nature of the time evolution. Also, one can distinguish the difference in the probability distributions from the shape of the bumps. The non-uniform distribution of the bump in the lower branch indicates the super-Poissonian distribution, whereas a squeezed bump is seen in the upper branches, indicating a squeezed state with the sub-Poissonian distribution.
	Unsurprisingly, the upper (lower) bump gets stronger (weaker) with the increment of the strength of the drive. However, the evolution process around the transition region is observed to be very slow, and therefore, a perfect steady state almost remains unachieved at the end of time evolution.
	
	For better visualization of the exact dynamical behavior of the evolution, we plot the evolution of the Wigner function through videographic representation in the \href{supplimentary.pdf}{Supplementary Material}, which clearly shows the bump of the Wigner function rotates in different directions in phase space, while evolving in different branches. Therefore, the splitting of the bump into two and moving in opposite directions in the transition region is also better understood there.

	We investigate the nonclassical dynamical behavior of the Kerr nonlinear system by plotting the time evolution of the second-order correlation function $(g^2(0))$ in Fig. \ref{fig:g20}. 
	It exhibits that the system intends evolving through bunching modes $(g^2(0) >1)$ for weaker drives, and anti-bunching modes $(g^2(0) <1)$ for stronger drives. 
	More importingly, the system starts evolving through bunching modes $(g^2(0) >1 )$ when the initial state remains at lower $(\alpha(0) = 0)$ or metastable $(\alpha(0)=[1.5,0.33\pi])$  branch. However, if the system starts from upper branch $(\alpha(0)=[2.5,-0.37\pi])$, it starts evolving through anti-bunching modes $(g^2(0) < 1 )$ . Also, following the Wigner function in Fig. 
	\ref{fig:wig_fn_1_6_16_41_101}
	(II and III) (also in \href{supplimentary.pdf}{Supplementary Material}), the evolution of $g^2(0)$ at Fig. \ref{fig:g20}(b) shows the system takes longer time to saturate. Throughout the evolution process, the non-classicality moreover exhibits remain highly valued when the system evolves through transition regions.

	Following the second-order correlation function, we analyzed the dynamical behavior of the system through the evolution of fidelity. Fig. \ref{fig:Fidality_E1_8_10_20} shows that the system moreover evolves through higher fidelity in the lower branch compared to the upper branch. 
	{ As the system evolves through a non-classical state when it steps into the transition region, due to superposition of two coherent branches; a rapid reduction of fidelity is observed when the system takes a quantum jump from one branch to another during evolution. However, when the system goes back to one among two stable branches, fidelity rises up ( Fig. \ref{fig:Fidality_E1_8_10_20}(a,d))}.
	{Since the fidelity measures how close the state is to its quasiclassical approximation, its remaining closer to unit value enables acceptability of the mean-field dynamics. However, a reduction from its unit value raises concern while accepting that approximation. Fig. \ref{fig:Fidality_E1_8_10_20}(a,d) also shows that while looking at the impacts of the initial state on fidelity measurements, we notice that the fidelity reduces more during the transition when the initial state belongs completely different (in a different branch) than the final state.}

	{A quantum state is identified to be Gaussian from the Gussianity of its characteristic function, which makes the Wigner function to be in Gaussian shape.
		Distortion of the bump of the Wigner function from its Gaussian shape in Fig. \ref{fig:wig_fn_1_6_16_41_101} 
		(also in \href{supplimentary.pdf}{Supplementary Material}) and the fluctuation of second-order correlation provokes to estimate of the non-Gaussianity of the Kerr-nonlinear system throughout evolution, which is quantified by the quantum relative entropy, given in Eq. \eqref{non_G_measure}.} It appears that the non-Gaussianity moreover follows the profile of fidelity. Fig. \ref{fig:nonG_E1_8_10_20} shows that the non-Gaussianity increases when the system jumps from one branch to another. The state tends to remain non-Gaussian throughout the transition region and approaches Gaussianity when it stabilizes to one of the branches. Following fidelity, it also shows the evolution of the system through the upper branch remains more non-Gaussian than the lower branch. 
	{Following fidelity, Fig. \ref{fig:wig_fn_1_6_16_41_101}(a,d) also shows 
	that the non-Gaussianity increases more during the transition when the initial state belongs completely different (in a different branch) than the final state.
}

	\section{CONCLUSION}\label{conclussion}

	{
		We have seen the TEBD numerical techniques can be used to study the dynamics of the Kerr nonlinear system, successfully, which brings better accuracy while determining the quantum fluctuations. 
		The estimation of the steady-state system field shows that the TEBD numerical simulation follows the quantum mechanical exact analytical solution, which exhibits the quantum jump occurring around the classically determined transition region. 
		The semi-classical solution exhibits bistability and determines branch values. The TEBD numerical result produces a superposition of two classical branches. The phenomenon, therefore, generates non-classical states, which have been confirmed by determining the second-order correlation function and Wigner function. Deviation from classical nature causes a reduction in the fidelity measurement and results in de-Gaussification of the state. While performing these characterizations, we notice that the system suffers a residual effect of the initial state, especially around the transition region. The phenomenon is witnessed as the earlier occurrence of the quantum jump of the system from bunching to the anti-bunching mode when it starts evolving from an initial state belonging to the upper branch.
		Wigner function exhibits bumps on the different locations of phase space for different steady states. The difference becomes prominent for the different branches of bistability. Analyzing the dynamical behavior of the system, we have seen that the bump rotates in the opposite direction when the system evolves through different branches. Around the transition region, the bump splits into two and rotates in opposite directions. This brings to witness the maximization of quantumness and therefore, de-Gaussification when the system suffers transition. We also observe that the evolution through the lower branch remains closer to the corresponding classical solution and therefore more Gaussian, than that of an upper branch. 	
		Based on the performance of the TEBD numerical technique, we conclude by saying that it is a promising platform to handle nonlinear open quantum systems. The method has been able to provide a quantum mechanical description of hysteresis experimentally observed before in nonlinear optical systems \cite{hysteresis_kerr_experimental_josa, hysteresis_kerr_experimental_PRL}
	}

	The Gaussian states are important for several functionals in quantum information, the study of non-Gaussianity could be useful in quantum communication and teleportation \cite{PRL_kerr_quantum_comm, PRA_kerr_quantum_comm}. Besides, it will also be useful for the phase control in switching systems \cite{Phase_control_switch}, especially in the determination of the acceptable range of the control drive around the transition region due to its nonclassical behavior. Also, the non-linear susceptibility has a strong control in the generation of TMSV, which has immense experimental applications. Besides, second-order nonlinear susceptibility, recently, third-order susceptibility has also been used in the generation of squeezed states in interferometers \cite{Kerr3_sqz}. More importantly, the quantum dynamics of Kerr-nonlinearity provides a strong base to study optical fields in nonlinear dispersive media \cite{superconducting_cavity_yale_2, photonic_crystal, waveguide}. Moreover, the analysis gives a useful background especially to study pulse propagation in nonlinear media and applicability in the fabrication of switching systems.

	\begin{acknowledgments}
		SA wishes to acknowledge the support of
		Ankush Kashiwar, Philippe Djorw\'e, Andr\'e Xuereb, and Abhishek Shukla for their technical advice while working on this project. The work has been supported by the European Union; Project number: 101065991 (acronym: SingletSQL)
	\end{acknowledgments}

	\appendix

	\section{TEBD FOR OPEN QUANTUM SYSTEM}  \label{TEBD_NUMERICAL_MODEL}

\begin{figure*}
		\centering
		\includegraphics[width=\linewidth]{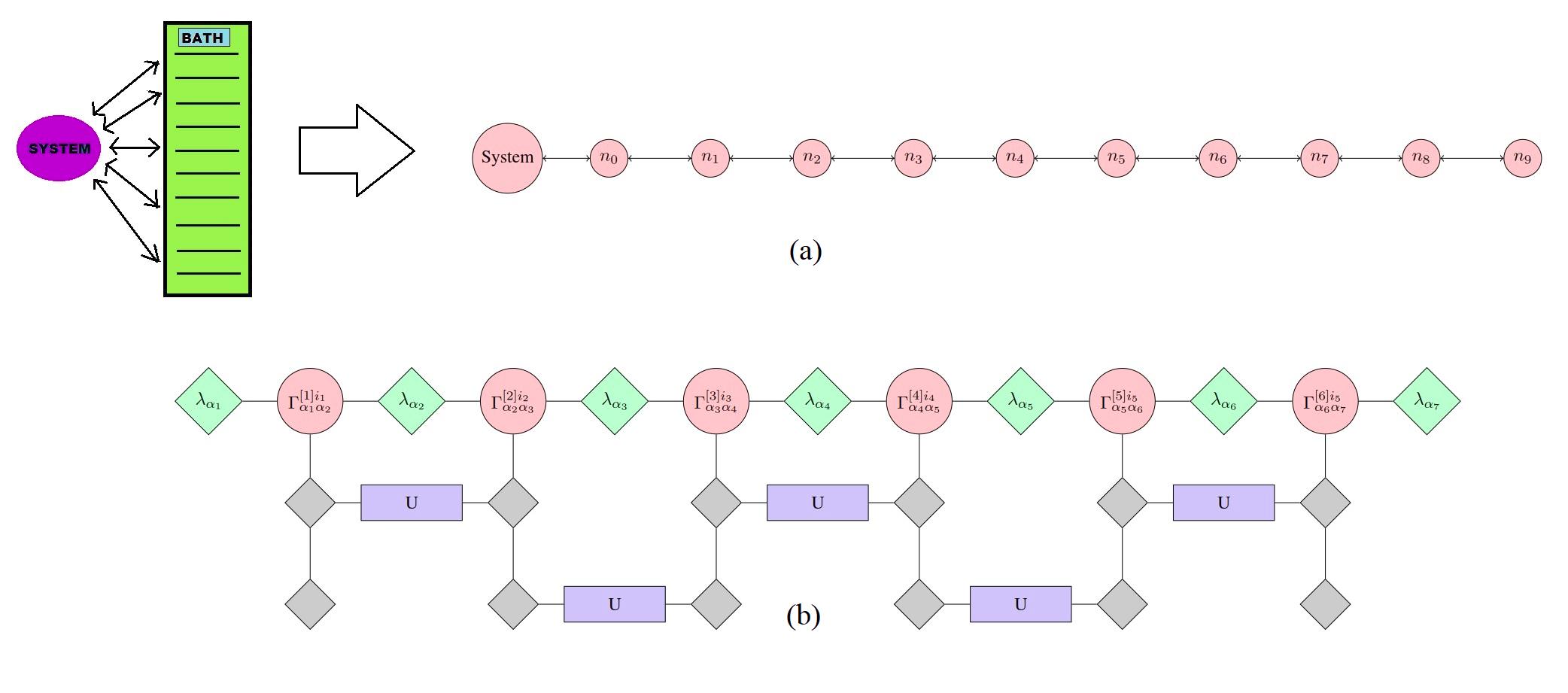}
		\caption{(a) Transformation of Hamiltonian from system/bath coupling model to semi-infinite chain model. (b) Diagrammatic expression of the real-time evolution operation on alternating pair}
		\label{fig:tebd}
	\end{figure*}

	The S/B coupling model of an open quantum system is defined as one of the systems being coupled to several modes of the bath. Therefore, it is required to transform the Hamiltonian bath operators to map to a semi-infinite lattice chain, which becomes useful to implement the TEBD numerical scheme for the simulation of the S/B coupling model. This is done through a unitary transformation: $ b_n = \int_{-{x_m}}^{x_m} U_n(x)d(x) \mathrm{d}x $. In this case, a normalized shifted Legendre polynomial has appeared to be a natural choice as the unitary operator for exhibiting spectral density in the form of Eq. \eqref{spctral_density}. The unitary operator $U_n(x) = \sqrt{\frac{(2n+1)}{2 x_{m}}} L_n(x/x_{m}) $, defined in the range of $x\in [-x_{m},x_{m}]$, satisfies orthogonality condition: $\int \mathrm{d}x U_n(x) U_m(x) = \delta_{n,m} $. Therefore, one obtains the transformed Hamiltonian as
	
	\begin{align}\label{chain_ham}
		{H}_{chain} & = H_S+ \eta' \left(a^\dagger b_0+ a b_0^\dagger\right)\\
		&+\lim_{N\to \infty} \left[ \sum_{n=0}^N \omega_n b_n^\dagger b_n+ \sum_{n=0}^{N-1} \eta_n \left(b_n^\dagger b_{n+1} + b_n b_{n+1}^\dagger\right)\right]\nonumber
	\end{align}
	
	where the coefficients are $\eta' = c_0\sqrt{2 \omega_c},\omega_n = 0$ and $,\eta_n = \omega_c \left(\frac{n+1}{\sqrt{(2n+1)(2n+3)}}\right)$. A schematic diagram of the transformation is presented in Fig. \ref{fig:tebd}(a). Such mappings have been used previously in Ref. \cite{openquanta_1dchain} for the simulation of open quantum systems which applies to spin-boson models \cite{openquanta_1dchain_spin_bosson} and biomolecular aggregates \cite{openquanta_1dchain_biomolecul}.

	Afterward, for the simulation of the chain using the TEBD algorithm, we express the state of the chain in terms of an MPS:
	
	\begin{align}
		|\Psi \rangle = &\sum_{{\alpha _{1},.,\alpha _{N +1}=0} }^{\chi} \sum_{ i_1...i_N=0}^{M} \lambda _{\alpha _{1}}^{[1]}\Gamma _{\alpha _{1}\alpha _{2}}^{[1]i_{2}} \lambda _{\alpha _{2}}^{[2]}\Gamma _{\alpha _{2}\alpha _{3}}^{[2]i_{3}}\cdot.\\
		&..\cdot \lambda _{\alpha _{N}}^{[N]}\Gamma _{\alpha _{N}}^{[N]i_{N}} \lambda _{\alpha _{N+1}}^{[N+1]}|{i_{1},i_{2},..,i_{N-1},i_{N}}\rangle\nonumber
	\end{align}
	
	The MPS state is obtained through the Schmidt decomposition of the pure state of N sites where $\chi$ is the Schmidt number and M is the dimension of local Hilbert space. This process expresses the coefficients of the state in the tensor product form and reduces the dimension from $ M^{N}$ to $N \times M \times \chi \times \chi $ where $\chi$ is the Schmidt number. The decomposition of a state is performed by singular value decomposition which generates bipartite splitting between two local Hilbert spaces in the chain. In this process, we only deal with $MN\chi^2 + (N+1)\chi$ terms. 
	Fig. \ref{fig:tebd} (b) shows the method of numerical simulation diagrammatically, for the real-time evolution, where we used second-order Suzuki Trotter (ST) expansion which resumes the unitary evolution operator as
	
	\begin{equation}
		U_{\delta t}= e^{-i\delta tH_{chain}}=e^{-iF\delta t/2}e^{-iG\delta t}e^{-iF\delta t/2}+O[\delta t^3]
	\end{equation}
	
	where, $ F=\sum_{i\ \mathrm{odd}}H_{chain}^{i,i+1}$ and $G=\sum_{i\ \mathrm{even}}H_{chain}^{i,i+1}$. The ST expansion minimizes the error in 3rd order of time steps by evolving the pairs of alternate sites.
	
	The simulation parameters are typically estimated by minimizing errors which appear mainly in two ways. Firstly, by modeling the S/B coupling Hamiltonian to a 1D chain, and secondly, by each step of simulation of the real-time evolution. The modeling error is contributed by the canonical transformation of S/B coupling to the 1D chain. In practice, we choose a model where the chain has a finite length because the number of modes of the bath is	finite, which causes the recurrence of the particle from the end of the chain. The recurrence time is dependent on the group velocity.  The recurrence time increases with the increment of the length of the chain, which happens	since the increment of the number of sites reduces the group velocity for the particle to travel. The increment of the cutoff frequency increases the group velocity, forcing the particle to	travel faster in the lattice, causing the reduction of the recurrence time. 
	Apart from this, there are two other major sources. One is the Suzuki-Trotter error is introduced by the finite size of the time step, and tends to concentrate on the overall phase for real-time evolution. The	accuracy of the simulation improved with the reduction time step. The other type of simulation error appears due to the finite sizes	of the time step and the truncation of the Hilbert space and Schmidt number. While truncating the Hilbert space and Schmidt number, one has to take care that the state shall be expressed with minimal error throughout evolution, i.e.	the set has to be complete.
	Previously, we discussed extensively a quantification of error while optimizing parameters in Ref.  \cite{Agasti_kerr, Agasti_OQS}.

	\section{ANALYTICAL REPRESENTATION OF STEADY STATE} \label{Analytical_Representation}
	
	The steady state of the Kerr non-linear system is determined in terms of the P function, by mapping the master equation to the Fokker-Planck equation \cite{Kerr_Drummond_Walls}. The P function is given by
	
	\begin{equation}
		P (\alpha,\beta) = \alpha^{p-2} \beta^{q-2} \exp(2\alpha\beta - \frac{E}{i\chi"\alpha} + \frac{E^*}{i\chi"\beta} ) ,
	\end{equation}
	
	which represents the state on a coherent basis as
	
	\begin{equation}
		\varrho = \int \frac{\vert \alpha \rangle \langle \beta^* \vert}{  \langle \beta^* \vert \alpha \rangle } P (\alpha,\beta) \mathrm{d} \mu(\alpha,\beta) ,
	\end{equation}
	
	where $\mu(\alpha,\beta)$ is the integration measure that can be chosen according to different classes of representations. The coherent basis is not orthogonal to each other, giving 
	
	\begin{equation}
		\langle \beta \vert \alpha \rangle = \exp ( \beta^* \alpha - |\alpha|^2/2 - |\beta|^2/2 ) .
	\end{equation}
	
	The normalization condition of the density matrix $(\varrho)$ imposes a normalization condition on the P function. Therefore, the normalization factor deduced as
	
	\begin{equation}
		\mathcal{N} =  \int  P (\alpha,\beta) \mathrm{d}  \mu(\alpha,\beta), 
	\end{equation}
	
	is expected to be unit valued $(\mathcal{N} = 1)$. To obtain the state on number basis, one has to expand the coherent states in terms of number basis:
	
	\begin{equation}
		| \alpha \rangle = \sum_n \frac{1}{ \sqrt{(n !) }} \exp (- |\alpha|^2/2 ) \alpha^n |n \rangle.
	\end{equation}
	
	This afterward expresses the state of the system as
	
	\begin{equation} \label{state_number_basis}
		\varrho = \int \sum_{n,m} \frac{  \alpha^n \beta^m }{ \sqrt{(n !m !) }}  e^{- \beta \alpha }  |n \rangle \langle m \vert P (\alpha,\beta)  \mathrm{d} \mu(\alpha,\beta) .
	\end{equation}
	
	To determine the matrix coefficients, one has to determine the moments of the state. One can define a generalized function that calculates all possible moments of the system, as 
	
	\begin{equation}
		G^{(m,n)} = \langle {a^\dagger }^m a^n \rangle = \frac{1}{\mathcal{N}}\int \beta^m \alpha^n P (\alpha,\beta) \mathrm{d} \mu(\alpha,\beta) ,
	\end{equation}
	
	which is determined by the Eq. \eqref{moment_calculating_generalized_function} \cite{Kerr_Drummond_Walls}. This eventually determines the state (Eq. \eqref{state_number_basis}) as
	
	\begin{equation}
		\varrho = \sum_{n,m} \frac{1}{ \sqrt{(n !m !) }} \sum_r \frac{ (-1)^r}{r!} G^{(m+r,n+r)} |n \rangle \langle m \vert .
	\end{equation}

	Another way of expressing the state in a coherent basis is through the Wigner function, for which one needs to define the characteristic function:
	
	\begin{equation}
		\chi (\eta) = \chi_N (\eta) e^{(-\frac{1}{2} |\eta|^2)}
	\end{equation}
	
	where $\chi_N (\eta)$ is the normally ordered characteristic function, defined by
	
	\begin{equation} \label{norm_char_func}
		\chi_N (\eta) = \text{Tr} \left[ \varrho e^{\eta a^\dagger} e^{-\eta^* a}\right] .
	\end{equation}
	
	The Wigner function is defined as
	
	\begin{equation}
		W(\zeta) = \frac{1}{\pi^2} \int e^{( \eta^*\zeta - \eta\zeta^*)} \chi (\eta) \mathrm{d}^2\eta.
	\end{equation}
	
	Substituting \eqref {state_number_basis} into Eq. \eqref{norm_char_func}, one can rewrite the Wigner function as
	
	\begin{equation}\label{Wig_fn_intermidiate}
		W(\zeta) = \frac{1}{\pi^2} \int P(\alpha,\beta) e^{\eta(\beta-\zeta^*) }  e^{-\eta^*(\alpha-\zeta) } e^{-\frac{1}{2} |\eta|^2} \mathrm{d}^2\eta \, \mathrm{d}\mu(\alpha,\beta) 
	\end{equation}
	
	One can evaluate the integral using the identity

	\begin{equation}
		\frac{1}{\pi} \int \mathrm{d}^2\eta \, \exp \left[ -\lambda |\eta|^2 +\mu \eta +\nu \eta^* \right] = \frac{1}{\lambda} \exp \left[ \frac{\mu \nu}{\lambda} \right]
	\end{equation}
	
	Replacing $\lambda = \frac{1}{2}, \, \mu = (\beta - \zeta^*), \, \nu = -(\alpha - \zeta)$, one rexpresses Eq. \eqref{Wig_fn_intermidiate} as
	
	\begin{equation}
		W(\zeta) = \frac{2}{\pi} \int P(\alpha,\beta) e^{-2(\beta-\zeta^*) (\alpha-\zeta) }  \mathrm{d}\mu(\alpha,\beta)
	\end{equation}
	
	Expanding the exponentials in the Power series
	
	\begin{widetext}
		\begin{align*}
			e^{-2(\beta-\zeta^*) (\alpha-\zeta) } 
			=  e^{-2( |\zeta|^2 ) } \left( \sum_k \frac{(-2)^k \beta^k \alpha^k}{k!}\right) \left( \sum_n \frac{(2)^n {\zeta^*}^n \alpha^n}{n!}\right) \left( \sum_m \frac{(2)^m \beta^m \zeta^m}{m!}\right),
		\end{align*}
	\end{widetext}
	
	and using the moment generating function in Eq. \eqref{moment_calculating_generalized_function}, 
	one determines the steady state Wigner function of the Coherent driven Kerr nonlinear system, which is given in Eq. \eqref{Wigner_function}.

	
	
	\nocite{*}
	
	\bibliography{apssamp}
	\bibliographystyle{apsrev4-2}

\end{document}